\documentclass[final,3p,times]{elsarticle}

\usepackage{amssymb}
\usepackage{arydshln} 
\usepackage{multirow}
\PassOptionsToPackage{hyphens}{url}\usepackage{hyperref}
\usepackage[table]{xcolor} 

\newcommand\BibTeX{{\rmfamily B\kern-.05em \textsc{i\kern-.025em b}\kern-.08em
T\kern-.1667em\lower.7ex\hbox{E}\kern-.125emX}}
\newcommand*\rot{\rotatebox{90}} 
\newcommand\gray{\cellcolor{gray!25}} 

\makeatletter
\def\ps@pprintTitle{%
 \let\@oddhead\@empty
 \let\@evenhead\@empty
 \def\@oddfoot{}%
 \let\@evenfoot\@oddfoot}
\makeatother

\begin{document}

\begin{frontmatter}

\title{Voting power of political parties in the Senate of Chile during the whole binomial system period: 1990-2017}

\author[label1]{Fabi\'an Riquelme}
\ead{fabian.riquelme@uv.cl}

\author[label2]{Pablo Gonz\'alez-Cantergiani}
\ead{gonzalezcantergiani@gmail.com}

\author[label3]{Gabriel Godoy}
\ead{gabriel.godoy@usach.cl}

\address[label1]{Escuela de Ingenier\'ia Civil Inform\'atica, Universidad de Valpara\'iso, Chile}
\address[label2]{Independent Researcher, Santiago de Chile, Chile}
\address[label3]{Departamento de Ingenier\'ia Inform\'atica, Universidad de Santiago de Chile, Chile}

\begin{abstract}
The binomial system is an electoral system unique in the world. It was used to elect the senators and deputies of Chile during 27 years, from the return of democracy in 1990 until 2017. In this paper we study the real voting power of the different political parties in the Senate of Chile during the whole binomial period. We not only consider the different legislative periods, but also any party changes between one period and the next. The real voting power is measured by considering power indices from cooperative game theory, which are based on the capability of the political parties to form winning coalitions. With this approach, we can do an analysis that goes beyond the simple count of parliamentary seats.
\end{abstract}


\end{frontmatter}


\section{Introduction}

Today, all Latin American countries have a \textit{congress} or \textit{assembly}, which represents the legislative power of their nation.\footnote{One exception could be Venezuela, where since 2016, the executive power, represented by Nicol\'as Maduro, has curbed the power of the legislative body, represented by the National Assembly of Venezuela.} These bodies can be classified as two types: the \textit{unicameral}, defined by a single Chamber of Deputies (Costa Rica, Cuba, El Salvador, Guatemala, Honduras, Nicaragua, Panama, and Venezuela), sometimes referred to as ``assemblymen'' (Ecuador) or ``congressmen'' (Peru); and the \textit{bicameral}, defined by a Chamber of Senators and another of Deputies (Argentina, Brasil, Chile, Haiti, Mexico, Paraguay, and Dominican Republic), which are sometimes also referred to as ``representatives'' (Colombia, Uruguay).

In Chile, legislative power lies with the National Congress, a bicameral body with a Senate and a Chamber of Deputies. Each parliamentary candidate represents a geographic sector of the country. The senators represent \textit{senatorial constituencies}, which normally agree with the country's regions, and the deputies represent electoral \textit{districts}, which are normally a group of neighboring communes. With regards to its structure, the Congress of Chile is no different most other bicameral bodies. However, its process of electing representatives between 1989 and 2017 distinguishes it from the rest of the world's legislative bodies.

Since return to democracy in 1989, parliamentary elections in Chile have taken place every four years, and as of this writing, the last one took place in November 2017. Over the last 28 years, these elections have been governed by the provisions of the current Political Constitution of 1980, established during the period of military dictatorship, and any subsequent amendments. Throughout this period, parliamentary elections were carried out through an electoral process known as the \textit{binomial system}, the only one of its kind in the world. The binomial system assigns two parliamentary seats to each electoral entity. For example, if two Candidates $X$ and $Y$ of the same political coalition have double the number of votes than another Candidate $Z$, then Candidates $X$ and $Y$ assume two seats, to the detriment of Candidate $Z$, even though the latter may have received more votes than $X$ or $Y$ when considered independently. This system has been very controversial. Its proponents believe that it maintains political stability, while its opponents point to problems of \textit{malapportionment} and argue that the system favors bipartisanship, at the expense of political mobility and the participation of independent candidates~\citep{Car06,Sia05,Nav01,RN05,MDM14}. According to \cite{Nav05}, at least until 2005, the political right coalition benefited the most from the binomial system. Therefore, their principal resistance to its replacement was not losing seats in Congress but rather the desire to maintain the status quo of a bipartisan system, supported by two broad political coalitions (the ``Alianza'' of the right, and the ``Concertaci\'on'' of the center-left) formed by a number of smaller political parties.

In November 2017, Chile faced a major change in the election of its senators and deputies. The binomial system was replaced by the well-known d'Hondt method, a highest-averages method that allocates seats in proportion to the votes obtained by each party-list. Moreover, the number of seats in Congress was increased and the distribution of seats among the country's districts was changed~\citep{Sen15,BCN15}. This will give rise to important changes in district~\citep{Alt14} and regional~\citep{RG17} representation and voting power. As a result, we can discuss the period of return to democracy, between 1989 and 2017, as a distinctive epoch in the history of the Chilean legislative system.

The main goal of this article is to study the real voting power of Chile's parties during the 1989-2017 Chilean legislative epoch. We restrict our analysis to the Senate, due to the considerable effort required to accounting for all changes in the political system and in politicians holding congressional seats (either because of a change in legislative period, illness, impeachment, or other extenuating circumstances). By ``real voting power'' we are referring to not only the number of senators belonging to each party, but also the power indices of the parties in each period and how different changes during distinct legislative periods affected the political trends of the Senate of Chile.

Power indices come from cooperative game theory, a multidisciplinary approach developed in the 1940s for the formal study of cooperation and conflict situations in rational decision-making processes~\citep{NM44}. Power indices are usually related to the capacity of each voter to form winning coalitions, which are a central structure in the study of voting systems~\citep{Rik62}. Various electoral systems in Europe and North America have been studied using this approach, with the determination of the real voting power of different politicians involved in each voting system as one of its main outcomes~\citep{TP08}. Despite this, the voting systems used in Latin America have rarely been studied with power indices. This study is presented as a crucial contribution to understanding how the Senate of Chile operates, complementing other approaches like those previously mentioned. These techniques can also be replicated by the study of the voting systems in other countries.

The rest of the paper is organized as follows. In the next section, we describe the different political parties that have been represented in the Senate of Chile between 1990 and 2017, as well as their political coalitions and political tendencies. Then we describe and analyze the changes in the political tendencies of the senators during that period. Taking into account this information and the different quorums associated with the kinds of law that are voted on, we define the different voting systems presented in the Senate during the binomial system epoch. These voting systems are represented as weighted voting systems, where the players are the different political parties and the political coalitions represented in the Senate. Next we apply power indices on these weighted voting systems, in order to measure the real voting power of each actor. Finally, we present our analysis, results, and the main conclusions of our work.

\section{Political Parties in the Senate of Chile (1990-2017)}

The first primary parliamentary elections following the return to democracy were held in December of 1989. The elected congressmen assumed their position in March of 1990. Since then, Congress was composed of 38 senators elected democratically, representing 19 senatorial constituencies. Until 2006, there were also between eight and eleven designated senators serving life terms. Although these senators do not represent any constituencies, they must be considered in this study because they also contribute to different political tendencies.

\begin{table*}[t]
\small\sf\centering
\caption{Political parties of Chile with representation in the Senate. The symbol ``/'' means change of name or replacement by an analogous body.}\label{tab:parties}
\begin{tabular}{llc}\\\hline
Political Coalition & Political Party & Years in the Senate\\\hline
  Concertaci\'on / Nueva Mayor\'ia
  & Movimiento Amplio Social (MAS) / Pa\'is   & 2008-2017 \\\cline{2-3}
  (Center-left)
  & Partido Socialista de Chile (PS)          & 1990-2017 \\\cline{2-3}
  & Partido Radical (PR) /                    &           \\
  & Socialdemocracia Chilena (SDCH) /         & 1990-2017 \\
  & Partido Radical Social Dem\'ocrata (PRSD) &           \\\cline{2-3}
  & Partido Por la Democracia (PPD)           & 1990-2017 \\\cline{2-3}
  & Democracia Cristiana (DC)                 & 1990-2017 \\\hline
  Out of pact
  & Democracia Regional Patag\'onica (DRP)    & 2013-2017 \\\cline{2-3}
  (Center-right)
  & Somos Ays\'en (SA)                          & 2014-2016 \\\cline{2-3}
  & Amplitud (Amp)                            & 2014-2017 \\\hline
  Alianza / Chile Vamos
  & ChilePrimero (CH1)                        & 2007-2010 \\\cline{2-3}
  (Right)
  & Partido Regionalista Independiente (PRI)  & 2009-2010 \\\cline{2-3}
  & Uni\'on Dem\'ocrata Independiente (UCC)       & 1994-2002 \\\cline{2-3}
  & Renovaci\'on Nacional (RN)                  & 1990-2017 \\\cline{2-3}
  & Uni\'on Dem\'ocrata Independiente (UDI)       & 1990-2017 \\\hline
\end{tabular}
\end{table*}

Since 1989, only 13 political parties had held representation in the Senate (see Table~\ref{tab:parties}).\footnote{Hereafter, for simplicity,  we will refer to the political parties by their abbreviations.} Of them, seven parties are still in power after elections of 2017, while the other six have only briefly held seats in the Senate. Three of these six parties (SA, CH1, UCC) have dissolved before 2017, while the other (PRI) has not had representation in Congress since 2010. If we consider that there are more than 30 political parties currently in Chile, and many other parties have been created and dissolved between 1989 and 2017, we can conclude that the parties represented in the Senate are only a small fraction of the total number of parties in Chile. However, almost all the largest and most traditional parties of the country have always held representation in the Senate, with the exception of the ``Partido Comunista'' (PC) which until 2017 held 6 of 120 seats in the Chamber of Deputies.

Political parties such as PC, PS, PRSD and DC are considered traditional parties, since they were formed prior to the 1973 Chilean coup. The first two political parties were banned during the military dictatorship and only resumed normal activities after the return of democracy. Around the second half of the 1980s, other parties were founded, such as PPD, created with the sole purpose of bringing down Augusto Pinochet in the 1988 national plebiscite and the two most popular right-wings parties, RN and UDI. Other parties, such as the right UCC and the center-right CH1 parties, had a short period of existence. The PRSD party was the result of a union of the PR and SDCH parties in August of 1994. In the last few years, new political parties have been created as well. These include Amplitud (right), MAS (left), and others focused on solving the problem of centralism, in particular DRP (center-right) and PRI (first center-right, and then right). MAS's only representation in the Senate was their founder, Alejandro Navarro, who later renounced MAS and created the Pa\'is party. For this reason, these two parties are considered together.

During the whole binomial system period, there had been strong bipartisanship in Chile. In fact, most of the parties in Congress traditionally collaborated in one of the two main political coalitions. The center, center-left and left coalition was originally called \textit{Concertaci\'on de Partidos por la Democracia}, until 2013 when two political parties (PC and MAS) joined and changed its name to \textit{Nueva Mayor\'ia}. On the other hand, the center-right and right coalition is called \textit{Chile Vamos}, although it has had other names including \textit{Alianza}, \textit{Coalici\'on por el Cambio}, and \textit{Uni\'on por el Progreso de Chile}, among others. The main parties and founders of this coalition are UDI and RN. To facilitate the naming conventions in this paper, we will refer to these coalitions as ``Concertaci\'on'' and ``Alianza'', respectively, the names most used during the whole period.

\begin{figure}[t]
\centering
\includegraphics[width=0.7\textwidth]{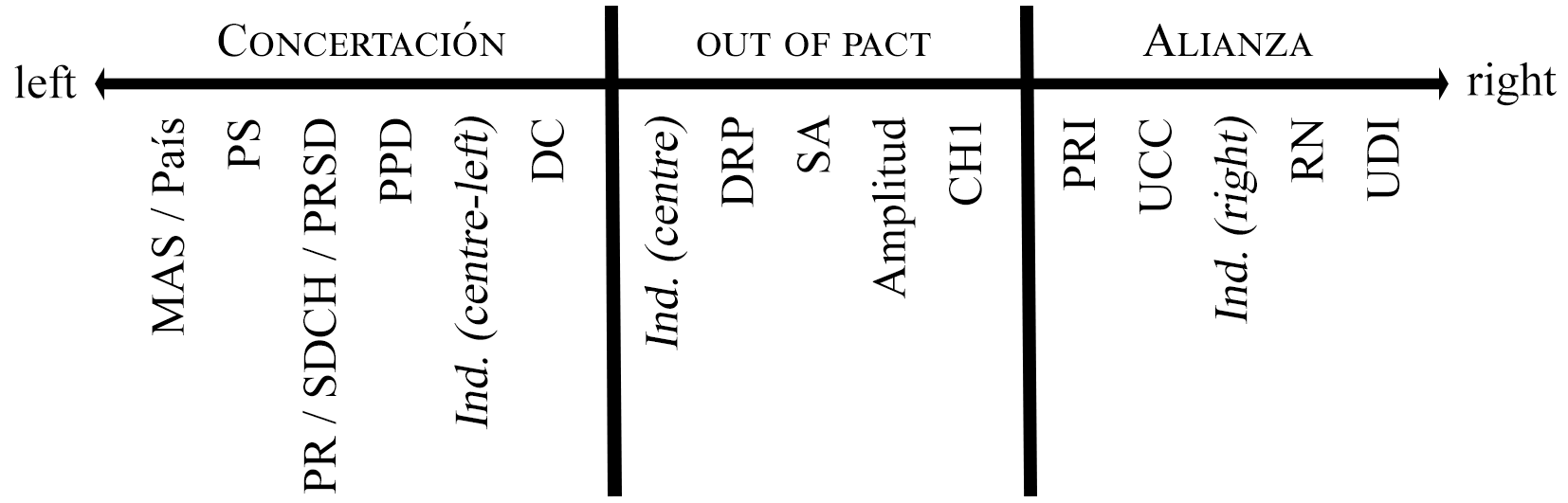}
\caption{Political parties with Senate representation from 1990 to 2017, grouped by their political coalition and ordered by their place on the political spectrum from left to right.}
\label{fig:parties}
\end{figure}

Figure~\ref{fig:parties} shows the political parties since 1990 with representation in the Senate of Chile, sorted by the political spectrum, from left to right. The figure also includes the independent parties and the parties that did not belong to a coalition. Note that the independent right parties always belonged to the Alianza. The DC party, although is considered a right-wing party in many countries, played an important role within the Concertaci\'on. The CH1 party belonged to the Alianza between May 2009 and January 2012, however only had a senator in office for less than one year. After its dissolution in January 2013, some of its members remained as center-right independents, while another group founded the {\em Partido Liberal} center party without any ties to the main coalitions.

For the last presidential elections of November 2017, the DC party rejected the idea of participating in the Nueva Mayor\'ia primaries, and supported their own candidate Carolina Goic, aside from the candidacy of Alejandro Guillier (Independent, pro-PRSD). This created a division within the Nueva Mayor\'ia coalition, which could result in a major rift in the future that coincides with the end of the binominal system.

\section{Historical Political Tendencies of Senators (1990-2017)}

In this section we present an exhaustive analysis about all the changes in political tendencies by senators since the beginning of the National Congress in 1990 until the end of the binominal system in 2017. The numbers of senators by party in each period and subperiod are described in Table~\ref{tab:subperiods}. The term \textit{period} describes the parliamentary period, beginning in March and ending in March four years later.  Changes in seats at the beginning of each period occur only from the results of the parliamentary elections of the previous year. We will use \textit{subperiods} to list all the small changes that occur in a period, due to any of the following factors:

\begin{table*}[!ht]
\small\sf\centering
\caption{Number of senators in each political party between 1990-2017. Dotted rows separate the different legislative periods. Gray squares mean that the party did not exist at that time.}\label{tab:subperiods}

{\footnotesize
\begin{tabular}{cccccccccccccccccc}\\\hline
Subperiod & \rot{MAS/Pa\'is} & \rot{PS} & \rot{PRSD} & \rot{PPD} & \rot{\textit{Ind. (center-left)}} & \rot{DC} & \rot{\textit{Ind. (center)}} & \rot{DRP} & \rot{SA} & \rot{Amplitud} & \rot{CH1} & \rot{PRI} & \rot{UCC} & \rot{\textit{Ind. (right)}} & \rot{RN} & \rot{UDI} & \rot{Total}\\\hline
  03/90-11/90 & \gray0 & 1 & 4 & 4 & 0 & 13 & 3 & \gray0 & \gray0 & \gray0 & \gray0 & \gray0 & \gray0 & 4 & 14 & 4 & 47\\
  11/90-04/91 & \gray0 & 4 & 4 & 1 & 0 & 13 & 3 & \gray0 & \gray0 & \gray0 & \gray0 & \gray0 & \gray0 & 3 & 14 & 4 & 46\\
  04/91-05/93 & \gray0 & 5 & 4 & 1 & 0 & 12 & 3 & \gray0 & \gray0 & \gray0 & \gray0 & \gray0 & \gray0 & 3 & 14 & 4 & 46\\
  05/93-03/94 & \gray0 & 5 & 4 & 1 & 0 & 12 & 3 & \gray0 & \gray0 & \gray0 & \gray0 & \gray0 & \gray0 & 3 & 15 & 3 & 46\\\cdashline{1-18}
  03/94-03/98 & \gray0 & 5 & 1 & 2 & 0 & 13 & 2 & \gray0 & \gray0 & \gray0 & \gray0 & \gray0 & 1 & 7 & 12 & 3 & 46\\\cdashline{1-18}
  03/98-06/98 & \gray0 & 4 & 2 & 2 & 0 & 15 & 2 & \gray0 & \gray0 & \gray0 & \gray0 & \gray0 & 1 & 6 & 7 & 9 & 48\\
  06/98-08/98 & \gray0 & 4 & 2 & 2 & 0 & 15 & 2 & \gray0 & \gray0 & \gray0 & \gray0 & \gray0 & 1 & 7 & 6 & 9 & 48\\
  08/98-03/00 & \gray0 & 4 & 2 & 2 & 0 & 15 & 2 & \gray0 & \gray0 & \gray0 & \gray0 & \gray0 & 0 & 7 & 6 & 9 & 47\\
  03/00-12/00 & \gray0 & 4 & 2 & 2 & 0 & 16 & 2 & \gray0 & \gray0 & \gray0 & \gray0 & \gray0 & 0 & 7 & 6 & 9 & 48\\
  12/00-01/02 & \gray0 & 4 & 2 & 2 & 0 & 16 & 2 & \gray0 & \gray0 & \gray0 & \gray0 & \gray0 & 0 & 6 & 6 & 10 & 48\\
  01/02-03/02 & \gray0 & 4 & 2 & 2 & 0 & 16 & 2 & \gray0 & \gray0 & \gray0 & \gray0 & \gray0 & 1 & 6 & 6 & 10 & 49\\\cdashline{1-18}
  03/02-07/02 & \gray0 & 5 & 2 & 3 & 0 & 14 & 2 & \gray0 & \gray0 & \gray0 & \gray0 & \gray0 & \gray{0} & 6 & 6 & 11 & 49\\
  07/02-12/02 & \gray0 & 5 & 2 & 3 & 0 & 14 & 2 & \gray0 & \gray0 & \gray0 & \gray0 & \gray0 & \gray{0} & 5 & 6 & 11 & 48\\
  12/02-03/03 & \gray0 & 5 & 2 & 2 & 1 & 14 & 2 & \gray0 & \gray0 & \gray0 & \gray0 & \gray0 & \gray{0} & 5 & 6 & 11 & 48\\
  03/03-03/05 & \gray0 & 5 & 2 & 2 & 1 & 14 & 2 & \gray0 & \gray0 & \gray0 & \gray0 & \gray0 & \gray{0} & 4 & 7 & 11 & 48\\
  03/05-06/05 & \gray0 & 5 & 3 & 2 & 0 & 14 & 2 & \gray0 & \gray0 & \gray0 & \gray0 & \gray0 & \gray{0} & 4 & 7 & 11 & 48\\
  06/05-03/06 & \gray0 & 5 & 4 & 2 & 0 & 13 & 2 & \gray0 & \gray0 & \gray0 & \gray0 & \gray0 & \gray{0} & 4 & 7 & 11 & 48\\\cdashline{1-18}
  03/06-11/06 & \gray0 & 8 & 3 & 3 & 0 & 6 & 1 & \gray0 & \gray0 & \gray0 & \gray0 & \gray0 & \gray{0} & 0 & 8 & 9 & 38\\
  11/06-05/07 & \gray0 & 8 & 3 & 2 & 0 & 6 & 2 & \gray0 & \gray0 & \gray0 & \gray0 & \gray0 & \gray{0} & 0 & 8 & 9 & 38\\
  05/07-11/07 & \gray0 & 8 & 3 & 2 & 0 & 6 & 1 & \gray0 & \gray0 & \gray0 & 1 & \gray0 & \gray{0} & 0 & 8 & 9 & 38\\
  11/07-12/07 & \gray0 & 8 & 3 & 2 & 0 & 6 & 1 & \gray0 & \gray0 & \gray0 & 1 & \gray0 & \gray{0} & 1 & 7 & 9 & 38\\
  12/07-11/08 & \gray0 & 8 & 3 & 2 & 0 & 5 & 1 & \gray0 & \gray0 & \gray0 & 1 & \gray0 & \gray{0} & 2 & 7 & 9 & 38\\
  11/08-01/09 & 1 & 7 & 3 & 2 & 0 & 5 & 1 & \gray0 & \gray0 & \gray0 & 1 & \gray0 & \gray{0} & 2 & 7 & 9 & 38\\
  01/09-07/09 & 1 & 7 & 3 & 2 & 0 & 5 & 1 & \gray0 & \gray0 & \gray0 & 1 & 1 & \gray{0} & 1 & 7 & 9 & 38\\
  07/09-03/10 & 1 & 7 & 3 & 1 & 1 & 5 & 1 & \gray0 & \gray0 & \gray0 & 1 & 1 & \gray{0} & 1 & 7 & 9 & 38\\\cdashline{1-18}
  03/10-02/12 & 1 & 5 & 1 & 4 & 0 & 9 & 1 & \gray0 & \gray0 & \gray0 & \gray0 & 0 & \gray{0} & 1 & 8 & 8 & 38\\
  02/12-12/13 & 1 & 5 & 1 & 4 & 0 & 9 & 1 & \gray0 & \gray0 & \gray0 & \gray0 & 0 & \gray{0} & 2 & 7 & 8 & 38\\
  12/13-01/14 & 1 & 5 & 1 & 4 & 0 & 9 & 0 & 2 & \gray0 & \gray0 & \gray0 & 0 & \gray{0} & 1 & 7 & 8 & 38\\
  01/14-03/14 & 1 & 5 & 1 & 4 & 0 & 9 & 0 & 2 & \gray0 & 1 & \gray0 & 0 & \gray{0} & 1 & 6 & 8 & 38\\\cdashline{1-18}
  03/14-10/14 & 1 & 6 & 1 & 6 & 0 & 7 & 0 & 2 & \gray0 & 1 & \gray0 & 0 & \gray{0} & 0 & 6 & 8 & 38\\
  10/14-07/16 & 1 & 6 & 1 & 6 & 0 & 7 & 0 & 1 & 1 & 1 & \gray0 & 0 & \gray{0} & 0 & 6 & 8 & 38\\
  07/16-11/16 & 1 & 6 & 1 & 6 & 0 & 7 & 0 & 1 & 1 & 1 & \gray0 & 0 & \gray{0} & 1 & 5 & 8 & 38\\
  11/16-01/17 & 1 & 5 & 1 & 6 & 1 & 7 & 0 & 1 & 0 & 2 & \gray0 & 0 & \gray{0} & 1 & 5 & 8 & 38\\
  01/17-04/17 & 1 & 5 & 1 & 6 & 1 & 7 & 0 & 1 & \gray0 & 2 & \gray0 & 0 & \gray{0} & 1 & 5 & 7 & 37\\
  07/17-11/17 & 1 & 5 & 1 & 6 & 1 & 7 & 0 & 1 & \gray0 & 2 & \gray0 & 0 & \gray{0} & 0 & 6 & 7 & 37\\
  11/17-12/17 & 1 & 5 & 1 & 6 & 1 & 7 & 0 & 1 & \gray0 & 2 & \gray0 & 0 & \gray{0} & 0 & 6 & 6 & 36\\\hline
\end{tabular}}
\end{table*}

\begin{enumerate}
    \item A senator changes his political tendency or renounces his party.
    \item A senator leaves his position to assume a ministry position in the government. His seat is replaced by another politician of the same coalition (but possibly from another party).
    \item A senator is removed from his position to be investigated or impeached for legal reasons.
    \item A senator dies during his time in office. In that case, the seat may be filled by another politician from the same party, or the seat will remain available until the next term if there is an upcoming election.
\end{enumerate}

The reasons for each change between one subperiod and another are detailed in Table~\ref{tab:reasons}.

\begin{table*}[t]
\small\sf\centering
\caption{Changes in political tendency in the Senate of Chile (1990-2017).}\label{tab:reasons}
{\scriptsize
\begin{tabular}{lp{100ex}}\\\hline
Subperiod & Reason for the change in the number of senators of each political party\\\hline
  03/90-11/90 & Beginning of the first legislative period\\
  11/90-04/91 & The designated senator and commander in chief of the Air Force, C\'esar Ruiz Danyau (Ind. right) dies. Ricardo N\'u\~nez, Jaime Gazmuri and Hern\'an Vodanovic (PPD) return to PS\\
  04/91-05/93 & Eduardo Frei (DC) resigns to begin his presidential candidacy and is replaced by Mar\'ia Elena Carrera (PS)\\
  05/93-03/94 & Jaime Guzm\'an (UDI) dies and is replaced by Miguel Otero (RN)\\\cdashline{1-2}
  03/94-03/98 & Change of legislative period\\\cdashline{1-2}
  03/98-06/98 & Change of legislative period\\
  06/98-08/98 & Francisco Prat (RN) resigns to his political party and remains Ind. (right)\\
  08/98-03/00 & Francisco Javier Err\'azuriz (UCC) is removed from his position for legal conflicts\\
  03/00-12/00 & Eduardo Frei (DC), ex-president of the Republic, assumes as senator for life\\
  12/00-01/02 & Francisco Prat (Ind. right) leaves his political independence and joins UDI\\
  01/02-03/02 & Francisco Javier Err\'azuriz (UCC) regains his parliamentary immunity\\\cdashline{1-2}
  03/02-07/02 & Change of legislative period\\
  07/02-12/02 & Augusto Pinochet (Ind. right) resigned as senator for life, with 86 years old\\
  12/02-03/03 & Nelson \'Avila (PPD) resigns his political party and remains Ind. (left)\\
  03/03-03/05 & Antonio Horvath (Ind. right) leaves his political independence and joins RN\\
  03/05-06/05 & Nelson \'Avila (Ind. left) leaves his political independence and joins PRSD\\
  06/05-03/06 & Jorge Lavandero (DC) is removed from his position for investigation of crimes related to child abuse, being replaced by Guillermo V\'asquez (PRSD)\\\cdashline{1-2}
  03/06-11/06 & Change of legislative period (end of senators for life)\\
  11/06-05/07 & Fernando Flores (PPD) unofficially resigns from his party and remains Ind. (center)\\
  05/07-11/07 & Fernando Flores (Ind.center) leaves his political independence and creates CH1 with others\\
  11/07-12/07 & Carlos Cantero (RN) resigns to his political party and remains Ind. (right)\\
  12/07-11/08 & Adolfo Zald\'ivar (DC) is expelled from his party for his opposition to the Concertaci\'on and remains Ind. (right)\\
  11/08-01/09 & Alejandro Navarro (PS) resigns from his political party and creates MAS\\
  01/09-07/09 & Adolfo Zald\'ivar leaves his political independence and joins to PRI as presidential candidate\\
  07/09-03/10 & Roberto Mu\~noz (PPD) resigns from his political party and remains Ind. (left)\\\cdashline{1-2}
  03/10-02/12 & Change of legislative period\\
  02/12-12/13 & Antonio Horvath leaves RN, behaving like Ind. (right), despite formalizing his resignation in December 2013\\
  12/13-01/14 & Carlos Bianchi (Ind. right) and Antonio Horvath (Ind. center) create DRP party\\
  01/14-03/14 & Lily P\'erez (RN) resigns from his political party and joins to Amplitud\\\cdashline{1-2}
  03/14-10/14 & Change of legislative period\\
  10/14-07/16 & Horvath (DRP) resigns to his political party and joins to SA, that in 2017 joins to FREVS\\
  07/16-11/16 & Manuel Jos\'e Ossand\'on (RN) resigns to his political party and remains Ind. (right). In June, Alejandro Navarro resigns to MAS and creates PAIS\\
  11/16-01/17 & Antonio Horvath (SA) resigns from his party and joins to Sentido Futuro's coalition, remaining as Ind. (Amplitud). Fulvio Rossi (PS) resigns from his party and remains Ind. (left)\\
  01/17-04/17 & Jaime Orpis (UDI) is removed from his position for legal conflicts and resigns from his party\\
  07/17-11/17 & Manuel Jos\'e Ossand\'on come back to RN\\
  11/17-12/17 & Iv\'an Moreira (UDI) is removed from his position for legal conflicts\\\hline
\end{tabular}}
\end{table*}

\subsection{Independent Senators}

Table~\ref{tab:subperiods} includes senators who do not belong to any political party. A politician can be an independent, with no affiliation to a party but in a political coalition. In fact, the independent center-left senators have traditionally represented the Concertaci\'on, and those of the right have belonged to the Alianza. Only the center independent senators are considered out of both pacts. To determine the political tendency of the independent senators, we classify them by which party list they belonged to when they were elected and the political leaning of those parties.

There are some politicians elected as independents, especially during the first legislative period, but as soon they assumed office, they joined a political party. In those cases, we consider them as a regular member of a political party. Thus, senators are only listed as an independent if they maintained their independent affiliations for at least a few months after they assumed their position.

Additionally, there are a few senators that are not registered with a party, but whose views are entirely aligned with some party, like in the case of Pedro Araya (Ind. DC) and Alejandro Guillier (Ind. PRSD). Given that the votes of these politicians are completely aligned with those of a party, we count them as party politicians instead of independents.

Finally, the designated senators of the Armed Forces of Chile, who held relevant positions during the military dictatorship, were considered as right independents, while the non-military senators coming from either the Supreme Court or the Comptroller General of Chile, were considered center independents.

\subsection{Historical Analysis}
\label{subsec:history}

Table~\ref{tab:subperiods} and Table~\ref{tab:reasons} present historical events relevant to Chilean politics. The first two presidential periods after the return of democracy were the Patricio Aylwin (1990-1994) and Eduardo Frei (1994-2000) administrations, both of whom were DC members. Both presidents counted on a strong DC presence in parliament. Indeed, the DC party, together with the PRSD party, were the most stable parties with regards to changes in the political tendency of its members. During the first legislative period (03/1990-03/1994), several PPD candidates were elected to the Senate, but quickly returned to their original party, the renewed PS. Some right-wing independents did the same and later officially joined the UDI and RN parties. In contrast, the second legislative period (03/1994-03/1998) was the most stable period of the whole binomial system era in the Senate because there were no changes in political tendency.

In the beginning, RN led the Alianza in both chambers. However, from 1998 onwards, they began to cede many seats to UDI, who had begun to gather strength as a result of the nearly successful presidential candidacy of Joaqu\'in Lav\'in (UDI), although Ricardo Lagos (PPD), a candidate backed by PS, ultimately defeated him in 1999. In part because of Lav\'in's popularity at the time, UDI positioned itself well in the parliamentary elections of 2002 as the main party of Alianza. In contrast, during the Lagos administration (2000-2006), DC, a party characterized by its stable membership, experienced a major crisis when one of their most influential senators, Adolfo Zald\'ivar, renounced the party. Nicknamed the ``colorines'' due to the red hair of Zald\'ivar, other senators and deputies followed suit to create PRI in 2006, leaving the Concertaci\'on and  moving closer to the center-right. DC has not been successful thus far in recovering their political power. Furthermore, during the first term of the socialist president Michelle Bachelet (2006-2010), which also coincided with the end of appointed senators and those with life terms, PS overpowered the DC party. This fifth legislative period (03/2006-03/2010) was the least stable for the Senate during the binomial system era, with a total of eight changes in party membership.

The 2009 parliamentary elections produced a strong increase of the left-wing in the National Congress and were the first elections since the return of democracy in which PC succeeded in electing deputies. Meanwhile, the next presidential elections saw the election of the first right-wing president, Sebasti\'an Pi\~nera (2010-2014) of RN, following the return of democracy. During the Pi\~nera's administration, both DC and PPD gained senate seats, to the detriment of PS. Michelle Bachelet's second term in office (2014-2018) brought a new increase in seats for PS, although less than that of 2006, at the expense of DC. In contrast, UDI, although strongly affected in the Chamber of Deputies, retained its seats in the Senate, along with the rest of the right-wing parties. Since 2015, Bachelet received low approval rates comparable to the Sebasti\'an Pi\~nera administration. This, together with other factors, such as the creation of a new political coalition called Frente Amplio that contains new parties independent of the Concertaci\'on and Alianza, the DC crisis, and especially the replacement of the binomial system, produced a notable transformation in the Chilean political paradigm after the parliamentary elections of November 2017.

With the binomial system, the only period in which the number of Alianza senators surpassed the number from Concertaci\'on was during the second legislative period (03/1994-03/1998), in which there were still several designated senators serving life terms, including Augusto Pinochet. This advantage of the Alianza was also accomplished in the subperiods between 12/2007 and 03/2010, during the fifth legislative period. In what follows we will see whether these differences in senators made a difference in the voting power of the different parties and coalitions.

\section{Voting Systems in the Senate of Chile}

Both chambers of the National Congress can propose, approve, modify and reject bills. In order to approve a law, there has to first be an agreement in the chamber where the motion was first presented, and then it is must be approved by the other chamber.

Table~\ref{tab:laws} shows the distinct types of legislation voted by the National Congress, paired with the quorum required to ratify, modify, or abolish them. The laws are ordered according to their respective normative power. In addition, there are treaties and decrees with less normative power since the executive is responsible for approving them~\citep{AC11}.

\begin{table*}[t]
\small\sf\centering
\caption{Types of laws subject to voting in the National Congress and their required motions.}\label{tab:laws}
\begin{tabular}{ll}\\\hline
Kind of law & Motion to be passed, modified or derogated\\\hline
  Constitutional reform & Depending on the Constitution's Chapter:\\
                        & 2/3 of parliamentarians in exercise\\
                        & 3/5 of parliamentarians in exercise\\
  Constitutional interpretation & 3/5 of parliamentarians in exercise\\
  Constitutional organic & 4/7 of parliamentarians in exercise\\
  Qualified quorum & Absolute majority of parliamentarians in exercise\\
  Common or ordinary & Simple majority of parliamentarians in the session\\\hline
\end{tabular}
\end{table*}

We will now consider the fluctuations in voting power within the Chilean Senate. Rather than focusing on the specific results of the voting, we look at the \textit{potential} power of each party. Hence, we let voting systems with abstentions as future work. This leaves out the study of common laws, which are dependent on the number of session attendees. Therefore, our study considers constitutional reform laws, constitutional interpretation laws, constitutional organic laws, and qualified quorum laws. For voting systems with abstentions, the interested writer can see~\citet{FZ03}.

\subsection{Voting Systems}
\label{subsec:VS}

A \textit{voting system} is given by a set of voters and a quorum necessary to carry out a motion, e.g., the passing of a bill. A \textit{winning coalition} is a group of voters with enough ``yes'' votes to achieve the necessary quorum. A \textit{losing coalition} is any coalition that is not a winning one, i.e. any group of voters that do not meet the required quorum. A \textit{minimal winning coalition} is a winning coalition that would become a losing one if any of the votes changed to a ``no.''

Note that all possible coalitions must be either a winning or losing one. This notion of ``coalition,'' i.e. a group of voters, should not be confused with a \textit{political coalition}, i.e. a group of political parties associated with a political tendency. A ``coalition'' (winning or losing) could contain voters from many political tendencies. It is also worth noting that given a winning coalition, if one or more voters change from a ``no'' to a ``yes,'' the new coalition would still be a winning one, and if voters left a losing coalition, it would still be considered a losing one. This property is known as the \textit{monotonicity} in voting systems.

In this work, voters are represented by political parties with representation in the Senate. Each voter or political party has a weight given by the number of sitting senators who belong to the party. This means that the winning coalitions are a group of political parties, in which the number of senators belonging to those parties reach the required quorum. Note that the quorum changes according to the type of law being voted on (see Table~\ref{tab:laws}).

The voting systems that use weights for each voter are called \textit{weighted voting systems}~\citep{TP08}, 
and can be formally represented by vectors $[q;w_1,w_2,\ldots, w_n]$, where $q$ is the required quorum of the system, depending on the kind of law, and for each voter or political party $i\in\{1,\ldots,n\}$, $w_i$ is the number of senators who belong to the party in that period or subperiod.

The voting power of a political party is not directly associated with the number of active senators that belong to the party. For instance, consider the hypothetical case of three political parties elected, each with 18, 10, and 10 senators, respectively. Then suppose they want to vote a law with qualified quorum  (see Table~\ref{tab:laws}). In this case, to approve the law 20 votes are required. To calculate the voting power we assume that all the senators of a political party vote in the same way (something not unusual in the National Congress). Then the system can be represented as a weighted voting system $[20;18,10,10]$. Note that, as $w_1+w_2=w_1+w_3=18+10\geq 20$ and $w_2+w_3=10+10\geq 20$, then for the law to be approved, it is necessary that at least two of the three parties vote in favor. Therefore, every group of at least two parties is a winning coalition, while any isolated party forms a losing coalition. Since no party can approve the project unless it has the support of at least one other party, then the three parties share the same voting power, even though the first party has almost twice of representatives than the other two.
The previous exercise shows the need to use more expressive tools to measure the real voting power. For this we used in Section~\ref{subsec:powerindices} the \textit{power indices} coming from cooperative game theory and voting theory~\citep{TZ99,TP08}.

In order to analyze the political parties in Chile between 1990 and 2017, we must consider 144 weighted voting systems, which are obtained from the 36 periods and subperiods illustrated in Table~\ref{tab:subperiods}, and considering the four quorums shown in Table~\ref{tab:laws}, namely $\lfloor 2/3\rfloor+1$, $\lfloor 3/5\rfloor+1$, $\lfloor4/7\rfloor+1$, and $\lfloor1/2\rfloor+1$. Thus, the games to be analyzed are the following:

\begin{center}
\begin{tabular}[H]{ll}
$[q, 0, 1, 4, 4, 0, 13, 3, 0, 0, 0, 0, 0, 0, 4, 14,  4]$ &
$[q, 0, 8, 3, 2, 0,  6, 2, 0, 0, 0, 0, 0, 0, 0,  8,  9]$ \\
$[q, 0, 4, 4, 1, 0, 13, 3, 0, 0, 0, 0, 0, 0, 3, 14,  4]$ &
$[q, 0, 8, 3, 2, 0,  6, 1, 0, 0, 0, 1, 0, 0, 0,  8,  9]$ \\
$[q, 0, 5, 4, 1, 0, 12, 3, 0, 0, 0, 0, 0, 0, 3, 14,  4]$ &
$[q, 0, 8, 3, 2, 0,  6, 1, 0, 0, 0, 1, 0, 0, 1,  7,  9]$ \\
$[q, 0, 5, 4, 1, 0, 12, 3, 0, 0, 0, 0, 0, 0, 3, 15,  3]$ &
$[q, 0, 8, 3, 2, 0,  5, 1, 0, 0, 0, 1, 0, 0, 2,  7,  9]$ \\
$[q, 0, 5, 1, 2, 0, 13, 2, 0, 0, 0, 0, 0, 1, 7, 12,  3]$ &
$[q, 1, 7, 3, 2, 0,  5, 1, 0, 0, 0, 1, 0, 0, 2,  7,  9]$ \\
$[q, 0, 4, 2, 2, 0, 15, 2, 0, 0, 0, 0, 0, 1, 6,  7 \,\,\,,  9]$ &
$[q, 1, 7, 3, 2, 0,  5, 1, 0, 0, 0, 1, 1, 0, 1,  7,  9]$ \\
$[q, 0, 4, 2, 2, 0, 15, 2, 0, 0, 0, 0, 0, 1, 7,  6,\,\,\,  9]$ &
$[q, 1, 7, 3, 1, 1,  5, 1, 0, 0, 0, 1, 1, 0, 1,  7,  9]$ \\
$[q, 0, 4, 2, 2, 0, 15, 2, 0, 0, 0, 0, 0, 0, 7,  6,\,\,\,  9]$ &
$[q, 1, 5, 1, 4, 0,  9, 1, 0, 0, 0, 0, 0, 0, 1,  8,  8]$ \\
$[q, 0, 4, 2, 2, 0, 16, 2, 0, 0, 0, 0, 0, 0, 7,  6,\,\,\,  9]$ &
$[q, 1, 5, 1, 4, 0,  9, 1, 0, 0, 0, 0, 0, 0, 2,  7,  8]$ \\
$[q, 0, 4, 2, 2, 0, 16, 2, 0, 0, 0, 0, 0, 0, 6,  6, 10]$ &
$[q, 1, 5, 1, 4, 0,  9, 0, 2, 0, 0, 0, 0, 0, 1,  7,  8]$ \\
$[q, 0, 4, 2, 2, 0, 16, 2, 0, 0, 0, 0, 0, 1, 6,  6, 10]$ &
$[q, 1, 5, 1, 4, 0,  9, 0, 2, 0, 1, 0, 0, 0, 1,  6,  8]$ \\
$[q, 0, 5, 2, 3, 0, 14, 2, 0, 0, 0, 0, 0, 0, 6,  6, 11]$ &
$[q, 1, 6, 1, 6, 0,  7, 0, 2, 0, 1, 0, 0, 0, 0,  6,  8]$ \\
$[q, 0, 5, 2, 3, 0, 14, 2, 0, 0, 0, 0, 0, 0, 5,  6, 11]$ &
$[q, 1, 6, 1, 6, 0,  7, 0, 1, 1, 1, 0, 0, 0, 0,  6,  8]$ \\
$[q, 0, 5, 2, 2, 1, 14, 2, 0, 0, 0, 0, 0, 0, 5,  6, 11]$ &
$[q, 1, 6, 1, 6, 0,  7, 0, 1, 1, 1, 0, 0, 0, 1,  5,  8]$ \\
$[q, 0, 5, 2, 2, 1, 14, 2, 0, 0, 0, 0, 0, 0, 4,  7, 11]$ &
$[q, 1, 5, 1, 6, 1,  7, 0, 1, 0, 2, 0, 0, 0, 1,  5,  8]$ \\
$[q, 0, 5, 3, 2, 0, 14, 2, 0, 0, 0, 0, 0, 0, 4,  7, 11]$ &
$[q, 1, 5, 1, 6, 1,  7, 0, 1, 0, 2, 0, 0, 0, 1,  5,  7]$ \\
$[q, 0, 5, 4, 2, 0, 13, 2, 0, 0, 0, 0, 0, 0, 4,  7, 11]$ &
$[q, 1, 5, 1, 6, 1,  7, 0, 1, 0, 2, 0, 0, 0, 0,  6,  7]$ \\
$[q, 0, 8, 3, 3, 0,\,\,\, 6, 1, 0, 0, 0, 0, 0, 0, 0,  8,\,\,\, 9]$ &
$[q, 1, 5, 1, 6, 1,  7, 0, 1, 0, 2, 0, 0, 0, 0,  6,  6]$ \\
\end{tabular}
\end{center}
where quota $q$ may assume different values, depending on the game. For instance, for a game with 38 senators, the quotas per each type of law must be 26, 23, 22 and 20. The order of weights refer to the parties as in Figure~\ref{fig:parties}: MAS/Pa\'is, PS, PRSD, PPD, Ind. (center-left), DC, Ind. (center), DRP, Amplitud, CH1, PRI, UCC, Ind. (right), RN, and UDI.

In addition to the political parties, we also analyze the political coalitions. From the same periods, subperiods, and quorums, we obtain another 144 weighted voting systems, but only with three players. The first player represents the Concertaci\'on, the second one the independent senators of the center and center-right wing, and the third one the Alianza. The weights of each game are the sum of the members that belong to each coalition. The weighted voting games are the following:

\begin{center}
\begin{tabular}[H]{cccccc}
$[q, 22, 3, 22]$ & $[q, 23, 2, 23]$ & $[q, 24, 2, 22]$ & $[q, 19, 2, 17]$ & $[q, 18, 1, 19]$ & $[q, 21, 3, 14]$\\
$[q, 22, 3, 21]$ & $[q, 23, 2, 22]$ & $[q, 24, 2, 22]$ & $[q, 19, 1, 18]$ & $[q, 20, 1, 17]$ & $[q, 21, 3, 14]$\\
$[q, 22, 3, 21]$ & $[q, 24, 2, 22]$ & $[q, 24, 2, 22]$ & $[q, 19, 1, 18]$ & $[q, 20, 1, 17]$ & $[q, 21, 3, 14]$\\
$[q, 22, 3, 21]$ & $[q, 24, 2, 22]$ & $[q, 24, 2, 22]$ & $[q, 18, 1, 19]$ & $[q, 20, 2, 16]$ & $[q, 21, 3, 13]$\\
$[q, 21, 2, 23]$ & $[q, 24, 2, 23]$ & $[q, 24, 2, 22]$ & $[q, 18, 1, 19]$ & $[q, 20, 3, 15]$ & $[q, 21, 3, 13]$\\
$[q, 23, 2, 23]$ & $[q, 24, 2, 23]$ & $[q, 20, 1, 17]$ & $[q, 18, 1, 19]$ & $[q, 21, 3, 14]$ & $[q, 21, 3, 12]$\\
\end{tabular}
\end{center}

In this case, several consecutive vectors are repeated because some changes in senatorial militancy do not imply a change in a political coalition.

\subsection{Power indices}
\label{subsec:powerindices}

Given a voting system, power indices are measures that determine the relevance of the voters and their contribution within the system to form winning coalitions.

In this paper we shall use two classical power indices from the literature, used in the study of voting systems: the \textit{Shapley-Shubik index}~\citep{SS54}, based on the winning coalitions, and the \textit{Deegan-Packel index}~\citep{DP78}, based on the minimal winning coalitions. There exist other power indices, such as the \textit{Banzhaf index}~\citep{Ban65}, also known as Penrose-Banzhaf index~\citep{Pen46}, the \textit{Johnston index}~\cite{Joh78}, and the \textit{Holler-Packel index}~\cite{HP83}, also called \textit{Public Good Index}~\citep{Hol98}. However, for weighted voting systems, Banzhaf and Johnston are ordinally equivalent to Shapley-Shubik~\citep{FMP12}, while the Holler-Packel index returns practically the same results than Deegan-Packel~\citep{Das14}.

Given a weighted voting game of $n$ voters, let $X$ denote any coalition. Consider a coalition that begins as an empty coalition and is built by adding different voters sequentially one by one. We say that a voter is a \textit{pivot} if when it is added to the losing coalition that is being formed, the coalition becomes winning. Moreover, a voter is \textit{critical} in a coalition if that coalition is winning, but removing the voter, it becomes losing. Let us denote as $S_i$ the set of coalitions in which the voter $i$ is critical.

The \textit{Shapley-Shubik index} ($SS$) of a voter $i$ is based on the number of possible voter sequences in which that voter is a pivot. To normalize this measure, the above value is divided by the total number of sequences in which we can arrange the voters:

$$SS(i)=\sum_{X\in S_i}\frac{(n-|X|)!(|X|-1)!}{n!}$$

Let $m_1,\ldots,m_j$ be the cardinalities of all the minimal winning coalitions that contain $i$. The \textit{Deegan-Packel index} ($DP$) is given by: 
$$DP(i)=\frac{p_i}{p_1+p_2+\ldots+p_j}$$
where $p_i=1/m_i$, for $1\leq i\leq j$. 

\section{Analysis and Results}

In this section, we analyze the voting power of the parties and political coalitions in the Senate of Chile during the seven legislative periods of the binomial system, from 1990 to 2017.~\footnote{Supplementary data related to this article can be found at  \url{https://figshare.com/articles/Indices_de_Poder_-_Senado_de_Chile/5450011}.}

Each period and subperiod was analyzed using the four voting quorums described in Table~\ref{tab:laws}, resulting in 144 different weighted voting systems for political parties and another set for political coalitions (see Section~\ref{subsec:VS}). In order to compute the voting power of each system we used the Shapley-Shubik and Deegan-Packel indices, giving a total of 576 calculations.\footnote{The code, uploaded under the GNU General Public License v3.0 license, is available at this link: \url{https://github.com/vagnur/Power-Indices/tree/master}.}

The results were correlated using different methods, which allowed us to discover high correlations between variables that would be nearly impossible to find manually. Because the data set forms a non-parametric space, i.e., the data do not seem to follow a probability distribution based on a fixed set of parameters, we use the classical Spearman ($\rho$) and Kendall ($\tau$) correlation coefficients. The Spearman coefficient assesses how well the relationship between two variables can be described using a monotonic function. The Kendall coefficient measures the ordinal association between two measured quantities, i.e., it measures the similarity of the orderings of the data when ranked by each of the quantities. Both coefficients return values between -1 and 1, where -1 means a perfect inverse correlation, 0 an empty correlation, and 1 a perfect correlation. The computation of the correlation coefficients has an associated $p$-value. As usual, we consider the standard $0.05$ cutoff as significance level, so that the null hypothesis is rejected when the $p$-value is lower than 0.05. In what follows, we restrict our attention to those correlation results that are considered statistically significant, i.e., those with a $p$-value lower than 0.05.

The correlations ranges can be interpreted as shown in the Table~\ref{tab:corr_levels}. In simple terms, a high correlation between two variables means that when one variable increases (or decreases) then the other also does. Furthermore, a high inverse correlation means that when one variable increases (or decreases) the other variable decreases (or increases, respectively).

\begin{table*}[t]
\small\sf\centering
\caption{Correlation levels.}\label{tab:corr_levels}
\begin{tabular}{cl}\\\hline
Value & Correlation level\\\hline
  -1.0 to -0.8 & Very high inverse correlation\\
  -0.8 to -0.6 & High inverse correlation\\
  -0.6 to -0.4 & Moderate inverse correlation\\
  -0.4 to  0.4 & Low and very low correlation\\
\, 0.4 to  0.6 & Moderate correlation\\
\, 0.6 to  0.8 & High correlation\\
\, 0.8 to  1.0 & Very high correlation\\\hline
\end{tabular}
\end{table*}

\subsection{Political Parties}

\begin{figure}[!ht]
\centering
\includegraphics[height=0.8\textheight]{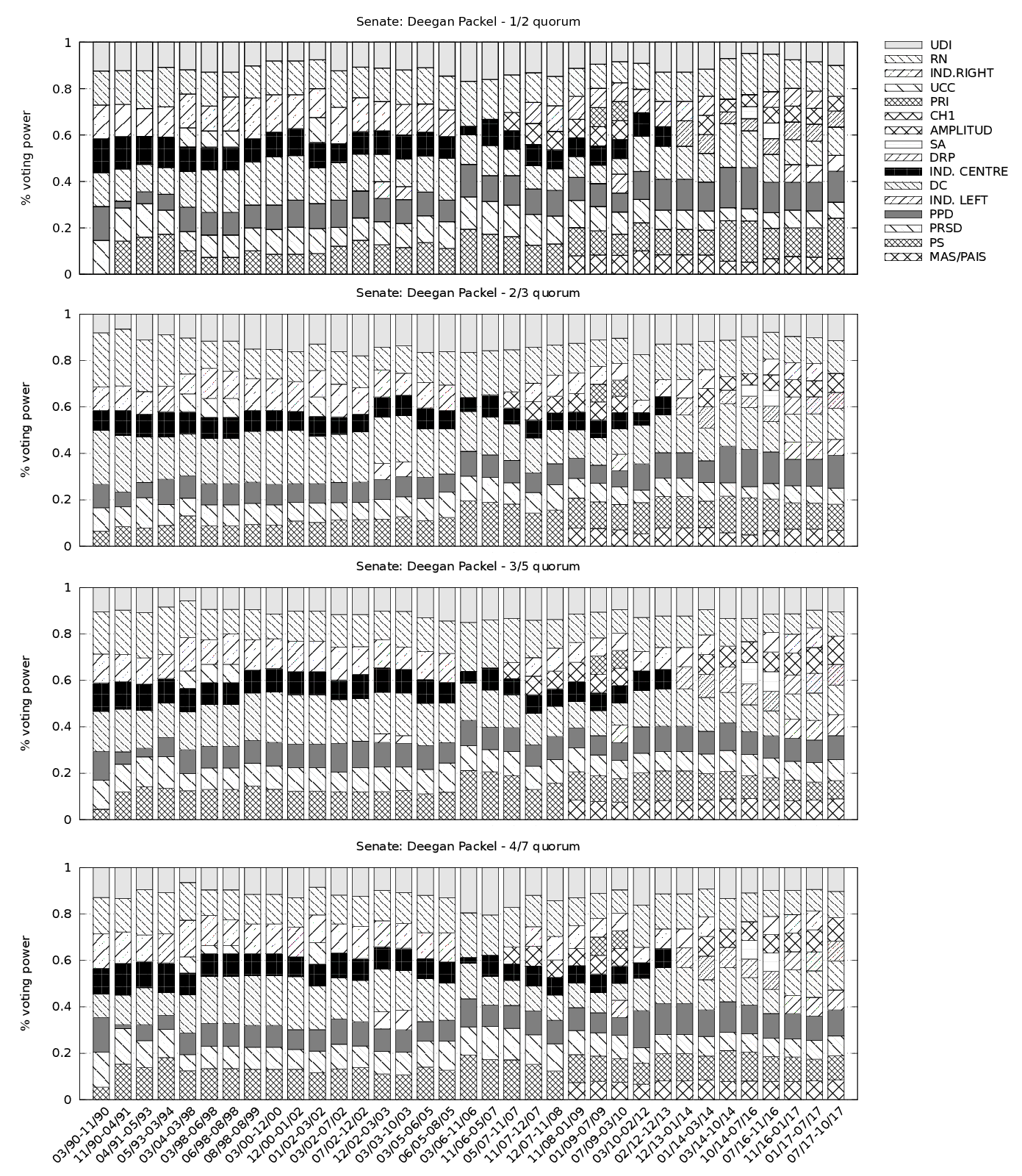}
\caption{Variation over time of the voting power for the different political parties, considering the different quorums, according to the Deegan-Packel index.}
\label{fig:graph-DP-parties}
\end{figure}

Figures~\ref{fig:graph-SS-parties} and~\ref{fig:graph-DP-parties} show the variation of the voting power of each party over time. For the Shapley-Shubik index, the following conclusions can be made:

\begin{figure}[!ht]
\centering
\includegraphics[height=0.8\textheight]{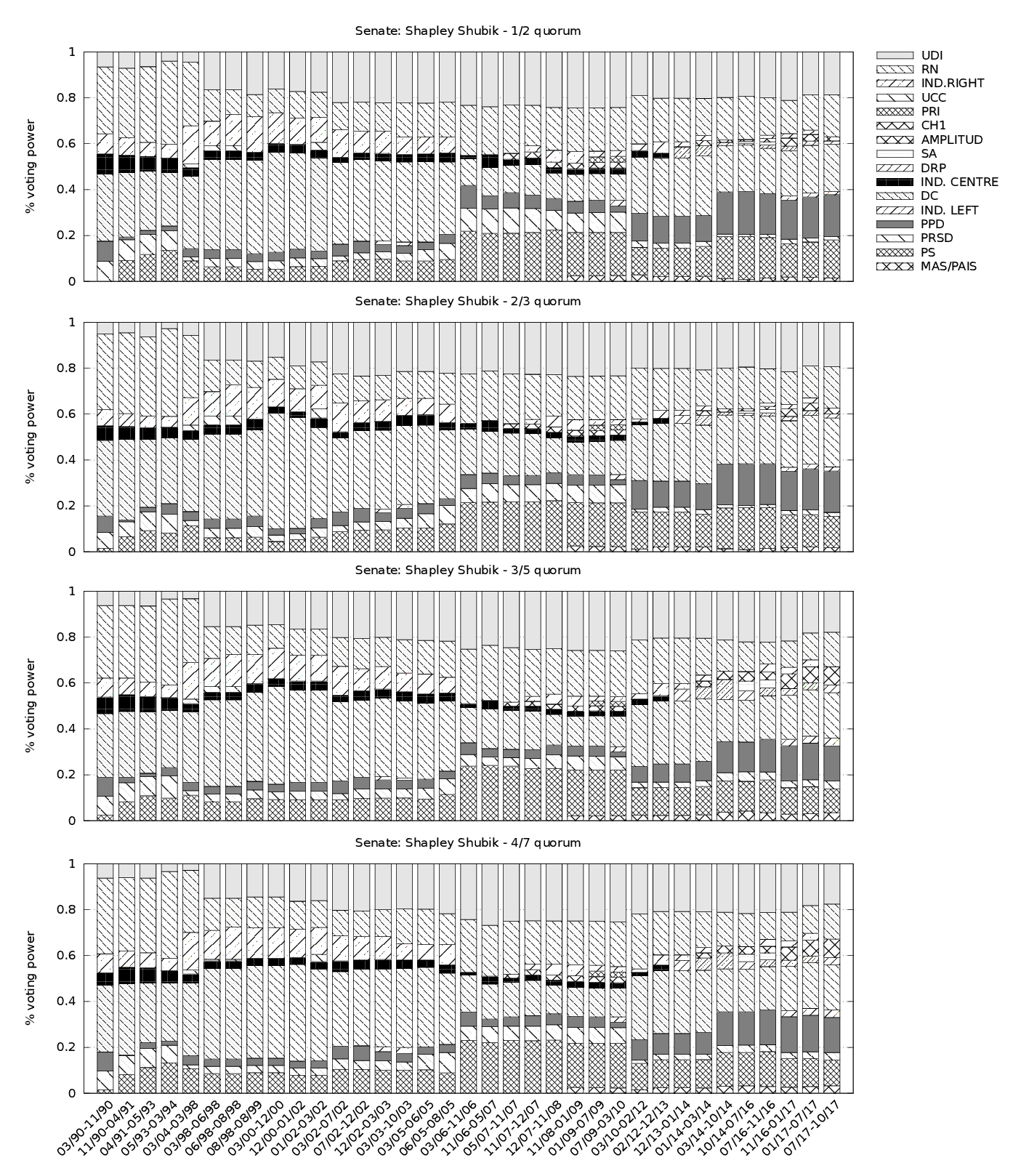}
\caption{Variation over time of the voting power for the different political parties, considering the different quorums, according to the Shapley-Shubik index.}
\label{fig:graph-SS-parties}
\end{figure}

\begin{itemize}
    \item The DC party and Ind. right have a high and very high Kendall correlation. This coincides with the fact that the expulsion of Adolfo Zald\'ivar from the DC party, in December 2007 ---which was one of the most important crises of that party--- did not affect significantly the historical voting power of the DC in the Senate.
    \item The PS and DC parties have a high and very high inverse correlation. This is especially noteworthy because it means that the two most relevant parties in the Concertaci\'on have traditionally disputed their voting power in the Senate. With the increase of seats of PS, in the fifth legislative period (03/2006-03/2010), the DC was the most affected party, while the decrease of seats of PS during the following periods resulted in a new increase of the power of the DC.
    \item The PS and Ind. right have a high and very high Kendall inverse correlation. This coincides with the high decrease of right-wing independent senators during the fifth legislative period. However, in this case, this decrease is not caused by an increase of voting power of the PS party, but by the end of the life senators, in which there were several right-wing independents.
    \item DRP and Amplitud parties have a high and very high correlation. This is not surprising, since both are center-right parties outside the pact, created in the penultimate legislative period. Moreover, they have only held between 1 and 2 senatorial seats. Therefore, their voting power have been always very similar.
    \item There is also a high correlation between MAS/PAIS and DRP, as well as between MAS/PAIS and Amplitud. However, this correlation is restricted to laws with 3/5 and 4/7 quorums. Although MAS/PAIS belongs to the Concertaci\'on, it has had only one representative in the Senate (Alejandro Navarro) and therefore its behavior is relatively similar to DRP and Amplitud. Another high correlation restricted to the same type of laws can be found between PS and UDI, but only for Kendall's correlation coefficient (for Spearman, both parties are not correlated).
\end{itemize}

For the same power index, we correlate the voting power obtained for each political party through the time with all the quorums. Thus we determine that the correlation between the quorums is high and very high for almost all political parties, except PRSD and UDI. This means that only for these two political parties, the Shapley-Shubik index returns significantly different values depending on the quorum considered.

A more detailed analysis of the results shows that the DC party leads in voting power during all legislative periods except the 1st, 5th and 7th ones. The first period was led by RN, leaving the DC in second place, except between March 1990 and April 1991, wherefore quorums of 1/3 and 1/2 both parties shared the first place. The fifth period was led entirely by the UDI, which from the second period remained in second place, moving RN to a third place, except between June 1998 and December 2000, when they reached the fourth place for certain quorums. The UDI resumed its hegemony at the beginning of the 7th period, but as of January 2017, after the resignation of the Senator Jaime Orpis due to legal conflicts, they had to share his leadership with the DC, yielding second place to the PPD. Before that, the PPD was in third place during all that period, first together with the PS and RN (until July 2016), then only with the PS (until November 2016), and finally alone (until January 2017). The PS is another party that has been well positioned within the Senate. It became the third prevailing party (after the DC and UDI) since April 1991. However, at the beginning of the second period, it fell to the fourth place, being surpassed by the right-wing independents, and even falling to the fifth position, in most of the quorums, until December 2000. In that month, with the signature of Francisco Prat for the UDI, the PS resumed the fourth indisputable place, after RN, until the end of the 4th period. Throughout the 5th period, under the first government of Socialist President Michelle Bachelet, the PS rose to second place, to the detriment of the DC, and remained alone between November 2007 and November 2008, to the detriment of RN. Finally, since the 6th period, the PS returned to yield its power to the DC, remaining in the third and fourth place, and being even surpassed by the PPD since November 2016, after the resignation of Senator Fulvio Rossi to the PS.

For the Deegan-Packel index, the results are quite different. This power index tends to enhance the voting power of smaller parties. In addition, it increases the variability in the rankings, and decreases the correlation between parties and quorums. Despite this, a very high correlation between DRP and Amplitud remained. Furthermore, only small parties behaved in a similar way regardless of the voting quorum. Regarding the DC, UDI, RN, PS, PPD, and PRSD parties, the results differ a lot depending on the quorum considered.

Due to the above, the hegemony, first of RN, and then of DC and UDI, is not so clear. For certain quorums, especially the lowest (1/3 and 1/2), the PS appears as the leader during some subperiods of the 1st period, during the first half of the 5th period, and since October 2014. For the same quorums, RN also reaches the first place, between the middle of the 4th period and the middle of the following period, at the beginning of the 6th period, and since October 2014. For this power index, it can be observed that changes in smaller parties may generate significant variations in the voting power of bigger parties. For example, the change in Horvath's membership from DRP to SA in October 2014, decreased the Deegan-Packel index of the DC in 15\% (from 0.184 to 0.156) and of the UDI in 28\% (from 0.09 to 0.065).

\subsection{Political Coalitions}

Just a few changes, of those listed in Table~\ref{tab:reasons}, affected the number of seats at political coalitions level. Nine changes affected the Alianza (3 positives and 6 negatives) and just three changes affected the Concertaci\'on (1 positive and 2 negative). Note that only one change affected both political coalitions, which was the expulsion of Adolfo Zald\'ivar from the DC in December 2007, for which he became right-wing independent.

In this case, the correlation results were unreliable due to the small amount of data, i.e., more technically, the correlation results produced $p$-values greater than 0.05. However, since the players are only three (Concertaci\'on, Alianza, and Out of pact) the results of the power indices can be analyzed more exhaustively.

\begin{figure}[t]
\centering
\includegraphics[height=0.5\textheight]{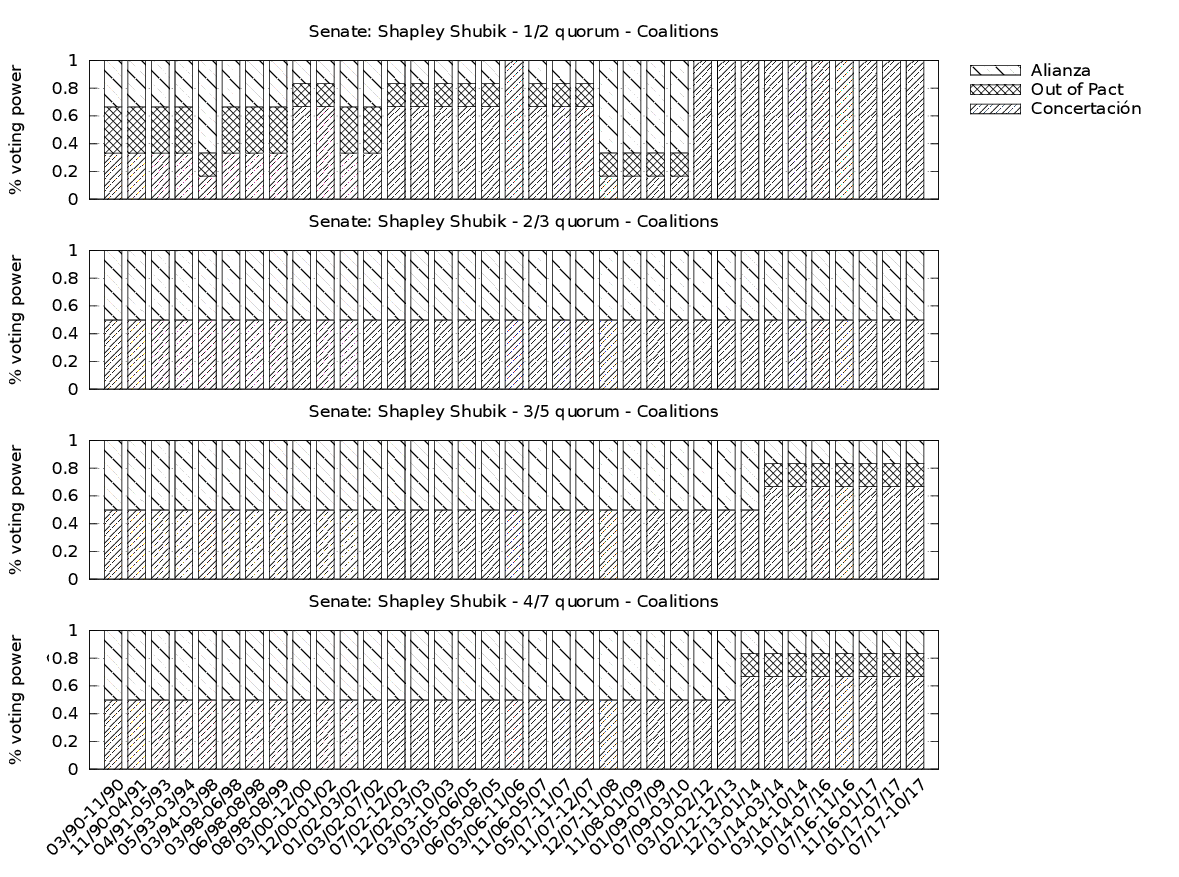}
\caption{Variation in the voting power of the political coalitions, for the different quorums, according to the Shapley-Shubik index.}
\label{fig:graph-SS-coal}
\end{figure}

The results for the Shapley-Shubik index are illustrated in Figure ~\ref{fig:graph-SS-coal}. Note that the voting power of the three players never completely differs, because in each subperiod at least two of them coincide. In addition, some changes in subperiods, namely, the death of Senator C\'esar Ruiz Danyau (right-wing independent), the impeachment of Francisco Javier Err\'azuriz (UCC), the founding of CH1 party by Fernando Flores, and the impeach of Jaime Orpis (UDI), did not generate any change in the voting power of the political coalitions.

For constitutional laws (2/3, 3/5, and 4/7 quorums), the Alianza and the Concertaci\'on maintained a full balance in their voting power for more than two decades. For constitutional reforms with 2/3 quorum, this balance remains unalterable. However, in December 2013, an apparently smaller event, such as the founding of the DRP party by Antonio Horvath (right-wind independent) and Carlos Bianchi, played in favor of the Concertaci\'on, for the case of constitutional organic laws (quorum 4/7). Since then, the Out of pact parties have emerged, with equal voting power than the Alianza, but far away from the power reached by the Concertaci\'on, which became the dominant political coalition. The effects of this Horvath move were reproduced identically in the following subperiod, but this time for constitutional laws with a 3/5 quorum, after Lily P\'erez left RN to enter Amplitud. Note that both changes came from the right-wing politics, and involved the emergence of small parties.

\begin{figure}[t]
\centering
\includegraphics[height=0.5\textheight]{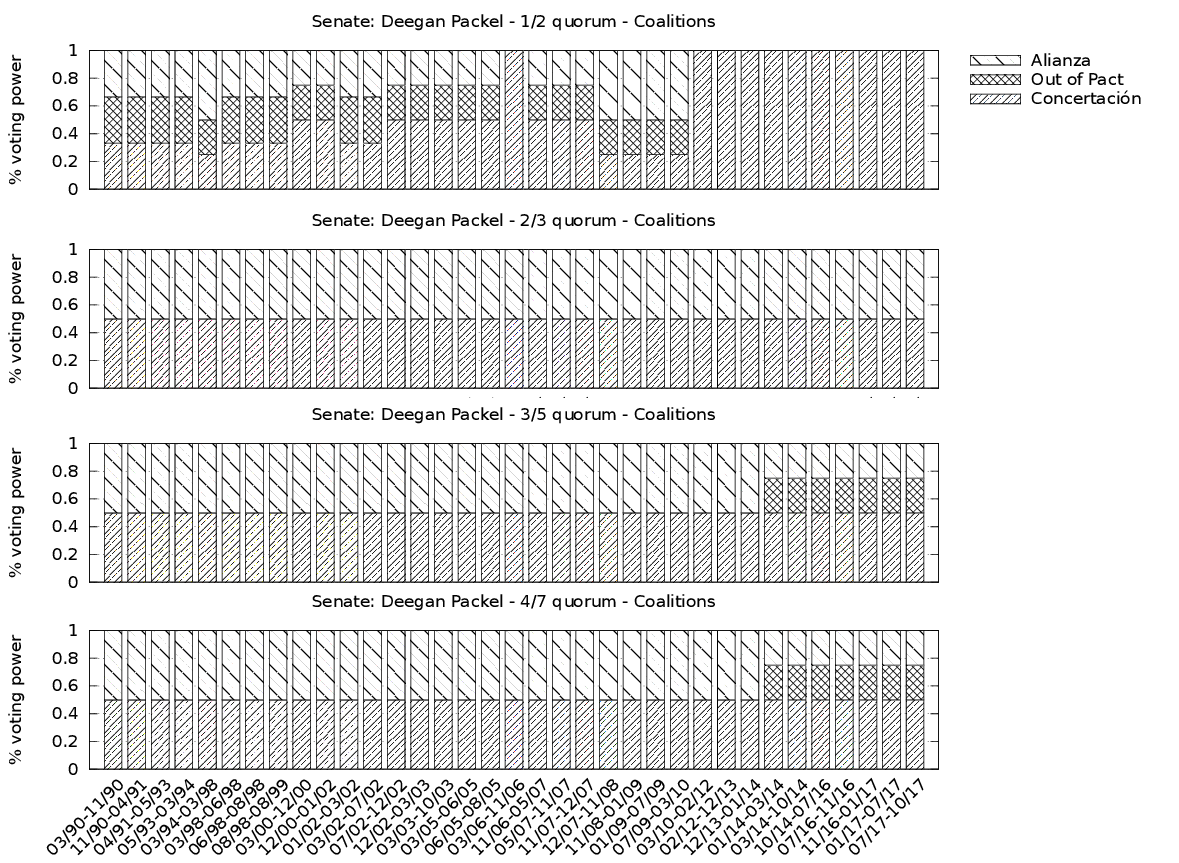}
\caption{Variation in the voting power of the political coalitions, for the different quorums, according to the Deegan-Packel index.}
\label{fig:graph-DP-coal}
\end{figure}

The case of qualified quorum laws (quorum 1/2) is the most complex, due to the more significant presence of the Out of pact. In this case, throughout the first period, started in March 1990, the three players have the same voting power. But in the second period in March 1994, the balance is broken, because the Alianza, with only two seats more than the Concertaci\'on, achieves more voting power than the Concertaci\'on and Out of pact together. In the March 1998 period the balance returns, but it is broken again by the entry of the ex-president Eduardo Frei (DC) as life senator. Note that this is the first period where voting power became in favor of the Concertaci\'on. This advantage continues until January 2002, when Err\'azuriz (UCC) regains its parliamentary jurisdiction. The balance is again broken in favor of the Concertaci\'on in July 2002, when Augusto Pinochet (right ind.) leaves his position of life senator. This new advantage of the Concertaci\'on continues until December 2007, and reaches its peak at the beginning of the fifth legislative period, in March 2006, when it achieved for the first time, during eight months, the absolute hegemony, which is lost with the retirement of Fernando Flores (PPD) and its approach to the independent center. In December 2007, after five years of leadership of the Concertaci\'on, the expulsion of Zald\'ivar from the DC and his switch to the Alianza reverse the roles, leaving the Alianza with more voting power than that of Concertaci\'on and Out of pact combined. This is only reversed with the following parliamentary elections, in March 2010. Since then, the Concertaci\'on has regained full voting power in the Senate.

Analyzing the presidential periods we can observe that for the constitutional laws, all the presidencies have had a balance of voting power between both political coalitions. There are two recent exceptions. The first exception was the case of Sebasti\'an Pi\~nera (2010-2014), who during his last months of government had to deal with the leadership of the opposition, the Concertaci\'on, for the laws with quorum 3/5 and 4/7. The second exception is the case of Michelle Bachelet (2014-2018), who has the leadership of the Concertaci\'on, the coalition to which she belongs, for the same quorum laws.

The case for the quorum laws (quorum 1/2) is more complex. The former president Patricio Aylwin (DC, 1990-1994) ruled with a full balance of power between the three players. By contrast, Eduardo Frei (DC, 1993-2000) had to deal during his first four years of government with a disadvantage of his political coalition, that only returned to the balance with the beginning of the next legislative period (March 1998). Ricardo Lagos (PPD, 2000-2006) was the first president that had for quorum laws a greater voting power of his political coalition, except for the brief period of balance between January and July 2002, when Err\'azuriz returned from his parliamentary impeachment. Bachelet's first government (PS, 2006-2010) began with full leadership of her political coalition, which remained a majority until December 2007. That month, Zald\'ivar's expulsion from the DC and his approach to the Alianza changed the situation, giving the leadership to the Alianza for the rest of her government. Sebasti\'an Pi\~nera's administration (2010-2014) has been by far the most affected by this type of law because throughout his presidency, the Concertaci\'on had an absolute voting power hegemony, which was maintained during Bachelet's second term (2014-2018).

Figure~\ref{fig:graph-DP-coal} shows that the Deegan-Packel has a similar behavior than Shapley-Shubik index, except for
a few exceptions. Note that in the constitutional organic laws, the change does not occur in December 2013 (with the founding of Horvath's DRP party), but happens instead in the next month, which found Lily P\'erez leaving RN for Amplitud. Note also that in all cases where Out of pact has some voting power, and one of the political coalitions holds the most power, then the leader's voting power is equal to the sum of the voting power of the other players. In this way, we observe again that the Deegan-Packel index brings the power of the players closer together. The latter allows that, except in those cases in which Concertaci\'on completely dominates the scene, minority players could work together to prevent the hegemony of the leader.

\section{Conclusions}

The binomial system is an electoral system unique in the world, used to elect the senators and deputies of the National Congress of Chile during 27 years, from the return of democracy in 1990 until 2017. In this article, we study the voting power of the different political parties and political coalitions in the Senate of Chile during the whole binomial period. We do not just consider the different legislative periods, but also the changes in the seats existing between one period and the next one. This provides a very detailed list of all relevant events that affected the parliamentarians in the last 27 years, resulting in 36 subperiods and more than 150 weighted voting games analyzed.

Using power indices coming from cooperative game theory we measured and analyzed the real voting power of each player. With this approach, we can do an analysis that goes beyond the simple count of parliamentary seats. This methodology allows us to find which parties lead the voting power in each subperiod and others remarkable findings, such as the case of DC party that led during all legislative periods except the 1st, 5th and 7th ones, according to the Shapley Shubik index. Using the Deegan-Packel index, we show how some smaller parties had noticeable power in the Senate. In addition, we found positives correlations like the one with DRP, Amplitud and MAS/PAIS; and inverse correlations like PS with the DC and Ind. right parties.

For the political coalitions, the analysis was more simple due to the inability to calculate correlations and also that just a few changes affected the number of seats in a coalition level. Besides that, some remarkable findings were made, like the subperiods where the Out of pact do not have any voting power, the large subperiods where the Alianza and Concertaci\'on have the same voting power (the totality of the 2/3 quorum and the majority of 3/5 quorum), and how the Concertaci\'on has a full voting power in the last nine periods for the 1/2 quorum laws. Furthermore, using the coalition's analysis, we could show which presidents had the Congress in his favor, and which ones, like the case of Sebastian Pi\~nera, had a strong opposition.

As a future work, we propose to compare the theoretical result of power indices with the actual results of the voting processes in the Senate. As a results, we could evaluate which parties, coalitions, or presidents took advantage according to their voting power. We also encourage others researchers to generate new and original analysis using the information with the relevant events that affect the numbers of seats in the Senate. Finally, it remains open the analysis of the Chamber of Deputies, as well as to use power indices variations that allow abstentions~\citep{FZ03}.

\section*{Acknowledgments}
This work was partially funded by project PMI USA1204.

We thank Alexa Jan for her comments and valuable proofreading of the article.

\end{document}